\documentclass[twocolumn,preprintnumbers,amsmath,amssymb]{revtex4}

\usepackage{amsmath}    % need for subequations
\usepackage{graphicx}   % for figures
\usepackage{color}

\begin{document}

\title{Using simple elastic bands to explain quantum mechanics: a conceptual review of two of Aert's machine-models}
\author{Massimiliano Sassoli de Bianchi}
\affiliation{Laboratorio di Autoricerca di Base, 6914 Carona, Switzerland}\date{\today}
\email{autoricerca@gmail.com}   %optional

\begin{abstract}

From the beginning of his research, the Belgian physicist Diederik Aerts has shown great creativity in inventing a number of concrete machine-models that have played an important role in the development of general mathematical and conceptual formalisms for the description of the physical reality. These models can also be used to demystify much of the strangeness in the behavior of quantum entities, by allowing to have a peek at what's going on -- in structural terms -- behind the ``quantum scenes,'' during a measurement. In this author's view, the importance of these machine-models, and of the approaches they have originated, have been so far seriously underappreciated by the physics community, despite their success in clarifying many challenges of quantum physics. To fill this gap, and encourage a greater number of researchers to take cognizance of the important work of so-called \emph{Geneva-Brussels school}, we describe and analyze in this paper two of Aerts' historical machine-models, whose operations are based on simple breakable elastic bands. The first one, called the \emph{spin quantum-machine}, is able to replicate the quantum probabilities associated with the spin measurement of a spin-$1/2$ entity. The second one, called the \emph{connected vessels of water model} (of which we shall present here an alternative version based on elastics) is able to violate Bell's inequality, as coincidence measurements on entangled states can do.\\

%\keywords{Quantum-machines \and Quantum probabilities \and Hidden measurements \and Spin \and Entanglement \and Bell's inequalities}

\end{abstract}

\maketitle

\section{Introduction}
\label{intro}

As is well known, Albert Einstein was convinced that God doesn't play dice with the universe. To that, Niels Bohr used to reply to stop telling God what s/he has to do with her/his dice. John. G. Cramer, in more recent times, added to this the following~\cite{Cramer}: ``If `God plays dice,' as Einstein (1932) has declined to believe, one would at least like a glimpse of the gaming apparatus that is in use.''

Having a glimpse of the apparatus means to \emph{understand} the nature of the game that is played behind the quantum scenes, and this in turn means \emph{to understand quantum mechanics}, something which most physicists believe is a hopeless goal, in accordance with Richard Feynman celebrated admonition~\cite{Feynman}: ``[...] that nobody understands quantum mechanics.''

But what does it mean to understand quantum mechanics? And, more generally, what does it mean to understand something in science? Different (more or less sophisticated) answers are of course possible, depending on the perspective which is adopted. But the majority of scientists will certainly agree that, roughly speaking, to understand in science means to build a theory with a sufficient explanatory power, which can suitably elucidate the observed phenomena and make confirmable predictions. Then, why quantum mechanics, which today has reached the level of a fully mathematized theory, and whose predictions have been confirmed experimentally to an extremely high degree of accuracy, is still considered a theory that nobody really understand, as Feynman used to proclaim? 

Again, the answer to this question will vary depending on the physicist that respond, but the majority will certainly say that although quantum theory describes and predicts with accuracy the behavior of quantum entities, the problem is that their behavior has nothing to do with that of classical macroscopic entities of our everyday experience. Therefore, we simply lack of concrete models that would allow us to visualize a quantum entity and understand, by analogy, its behavior, since we humans can only visualize models made of concrete objects, belonging to our ordinary spacetime theatre. 

Let us think for instance of an \emph{armilla} (also known as \emph{spherical astrolabe}): an ancient little model of the objects in the sky, consisting of a spherical framework of graduated metal circles (``armilla'' means ``circle'' in Latin), centered on Earth, representing the lines of celestial longitude and latitude, the ecliptic, and other astronomically important features. The armilla, which was invented by the Greek astronomer Eratosthenes (276 -194 BCE), was used, among other things, as a valid teaching tool, i.e., as a mean to conveniently visualize the structure and behavior of the astronomical solar system which, otherwise, because of its astronomical size, would have been rather difficult to visualize as a whole, and therefore to understand. 

Being the solar system an astronomical object, it is clear that any model of it will not be a real-size one, but a miniature. And  it is precisely because the armilla was a miniature, whose structure could be viewed as a whole, with a glance, that it had such a great explicative power and allowed to really understand the structure and behavior of the solar system (from an ancient geocentric point of view). 

The above is just to emphasize that the ability to create concrete, easily viewable models, is an important tool of the scientific enterprise, above all with regard to the possibility of fully understanding the functioning of the objects of its study. Now, if for astronomical entities models have to be miniatures, when it is about the representation of microscopic entities, models of course have to be larger than the original. As a paradigmatic example, we can consider the many different large scale atomic models that have been used in the course of history, to visualize and explore the atomic hypothesis. 

Let us think for instance of the very simple ``billiard ball'' atomic models of Democritus and Dalton, the more sophisticated Thomson's ``plum pudding'' model, or the Rutherford's ``armilla-like'' solar system model, subsequently modified by Bohr, by restricting the ``allowed'' orbits and energies for the electronic satellites.

This is probably one of the reasons why Bohr's atomic model, despite being totally obsolete, is sometimes still taught today as an introduction to quantum mechanics, as it has the merit of being somehow still viewable by the students, who have then the impression of understanding what's going on. But of course, Bohr's atomic model is already a ``crazy'' one, and Bohr himself was accustomed to say that the real question was to determine whether it was crazy enough to have a chance of being right! 

As we today all know, quantum theory, despite its weirdness, has certainly being confirmed to be right. But the price we have paid for this is, apparently, to renounce the possibility of understanding it. Indeed, what kind of model can we conceive today that would allow us to understand atoms without invoking magic images? (Like the one of an elevator which, in the case of Bohr's model, would be able to bring an electron to different floors, i.e., orbits, without ever sojourning in between of them).

To quote Heisenberg~\cite{Heisenberg}: ``The atom of modern physics can only be symbolized by a partial differential equation in an abstract multidimensional space. Only the experiment of an observer forces the atom to indicate a position, a color and a quantity of heat. All the qualities of the atom of modern physics are derived, it has no immediate and direct physical properties at all, i.e. every type of visual conception we might wish to design is, \emph{eo ipso}, faulty.''

The same of course can be said about the constituents of the atomic structures, the so-called \emph{elementary particles}, which in fact do not behave as particles, and can certainly not be \emph{understood} (i.e., visualized) as little tiny corpuscles moving around in the three-dimensional Euclidean space~\cite{Massimiliano, Massimiliano2}. So, how should we understand the reality of the microscopic quantum entities? And, more importantly, can we understand it? The majority of physicists believe that such an understanding is simply impossible, and this because we lack of simple models (apart from the abstract model of the mathematical theory) that would guide our intuition and allow us to understand the reasons for the observed quantum weirdness. But, is all this true? Do we really lack of simple concrete models that can help us to understand the behavior of elementary quantum entities?

In fact, and contrary to what is usually believed, these models exist, since many years now, and can certainly be used to convincingly visualize some of the strange behaviors of quantum entities, as they interact with the experimental contexts. They have been invented by the Belgian physicist Diederik Aerts, one of the founders of so-called \emph{Geneva-Brussels school on the foundations of physics}, and according to the present author their importance have been seriously underappreciated by the physics community, despite the role they have played in clarifying many of the conceptual challenges of quantum physics, and in facilitating the development of more advanced conceptual and mathematical formalisms that go beyond the quantum (Hilbertian) and classical (phase space) descriptions. 

Also, apart their interest in sponsoring new innovative approaches to understand the complexity of our reality -- like for instance the \emph{hidden-measurement approach}~\cite{Aerts4, Aerts4b, Aerts7, Aerts10} and \emph{creation-discovery view}~\cite{Aerts3, Aerts4, Aerts4b} -- these models can be advantageously used to reveal what could be a possible physical content of the theory, beyond its rather abstract and counterintuitive formalism. 

In this paper we shall present two of Aerts' historical quantum machine-models~\footnote{The term ``quantum machine'' might evoke in some readers the famous quantum machine of O'Connell~\cite{Connell}, whose logic is however very different from that of quantum machines invented by Aerts. Indeed, Aerts' quantum machines are conventional objects, whose quantum behavior is not a consequence of their internal coherence, but of the structure of the possibilities of actively experimenting with them, as a consequence of the peculiarities of the chosen experimental protocols.}, whose operations are based on simple breakable elastic bands. The first one is called the \emph{spin quantum machine}, or \emph{sphere model}~\cite{Aerts3, Aerts3-a}, and can easily replicate the quantum probabilities associated with the spin measurement of a spin-$1/2$ entity, for instance in a typical Stern-Gerlach experiment. The model allows to understand what possibly distinguish quantum from classical probabilities: if the latter correspond to a situation of lack of knowledge about the state of the system, the former may correspond to situations where there is full knowledge of the state of the entity, but maximum lack of knowledge about the interaction between the measurement apparatus and the entity.

The second machine-model that we shall describe is called the \emph{connected vessels of water} model~\cite{Aerts2, Aerts5, AertsBroekaert}. As its name indicates, the original model is constructed using vessels, tubes and water. However, we shall present here an alternative version of it, only using an elastic band, in accordance with the title of the present paper. The model is able to violate Bell's inequalities, as coincidence measurements on entangled states can do, and since everything in it happens under our eyes, part of the mystery of so-called ``spooky actions at a distance'' (i.e., EPR non-local correlations) will be revealed, and therefore explained: Bell's inequalities (and corresponding Bell's locality hypothesis) can only be violated when, during a coincidence experiment, correlations that weren't present before the experiment are literally created by and during the experiment itself.

\section{The spin quantum-machine model}
\label{Aert's spin quantum-machine model}

\subsection{The spin-$1/2$ quantum entity}
\label{The spin-$1/2$ quantum entity}

Before describing Aerts' spin quantum-machine (SQM), let us recall some of the basic properties of a spin-$1/2$ quantum entity. Its state $|\psi\rangle$ belongs to a 2-dimensional Hilbert space ${\cal H} = \mathbb{C}^2$, and can always be represented as a superposition of ``up'' and ``down'' spin states, relative to an a priori given direction $\hat{z}$:
\begin{equation}
\label{spin state}
|\psi\rangle = \alpha |+\rangle_{\hat{z}} + \beta |-\rangle_{\hat{z}}, 
\end{equation}
where $\alpha$ and $\beta$ are complex numbers obeying the normalization condition $|\alpha|^2 + |\beta|^2 =1$. 

If we denote by $\vec{S} = (S_{\hat{x}},S_{\hat{y}},S_{\hat{z}})$ the spin vector observable, its components are known to obey the commutation relation $[S_{\hat{x}},S_{\hat{y}}]=i\hbar S_{\hat{z}}$, and the equations: 
\begin{eqnarray}
\label{eigenvalue equation}
S_{\hat{z}}|\pm\rangle_{\hat{z}} &=& \pm\frac{\hbar}{2}|\pm\rangle_{\hat{z}},\\
\label{x-y component equation}
S_{\hat{x}}|\pm\rangle_{\hat{z}}&=&\frac{\hbar}{2}|\mp\rangle_{\hat{z}},\quad S_{\hat{y}}|\pm\rangle_{\hat{z}}=\pm i\frac{\hbar}{2}|\mp\rangle_{\hat{z}}.
\end{eqnarray}

Because of the normalization condition, one can always write (apart from a global phase factor with no physical meaning) $\alpha = \cos{\frac{\theta}{2}}\exp{-i\frac{\phi}{2}}$ and $\beta = \sin{\frac{\theta}{2}}\exp{i\frac{\phi}{2}}$, so that if $\theta$ and $\phi$ are taken to be the polar angles of a unit vector $\hat{v}$, then to each state vector $|\psi\rangle$ one can associate, bijectively, a unit vector $\hat{v}$, which can be written as $\hat{v}=(1, \theta,\phi)$ in polar coordinates, or as 
\begin{eqnarray}
\label{unit vector}
\hat{v}&=(\sin{\theta}\cos{\phi},\sin{\theta}\sin{\phi},\cos{\theta})\nonumber\\
&=(2 \Re \alpha^*\beta, 2 \Im \alpha^*\beta, |\alpha|^2-|\beta|^2)
\end{eqnarray}
in Cartesian coordinates.

According to the above one-to-one Pauli mapping, all spin states can be visualized as points on a 3-dimensional unit sphere, indexed by a specific unit vector (this is also known as a Poincar\'e sphere representation; see for instance the discussion in~\cite{Auletta}, pp. 26--28). More precisely, if $|\psi\rangle$ is given by (\ref{spin state}), and $\hat{v}$ is given by (\ref{unit vector}), then we have $|\psi\rangle \propto |+\rangle_{\hat{v}}$ (where the symbol ``$\propto$'' means here ``equals, up to a non zero overall phase factor''), and clearly $|+\rangle_{-\hat{v}}=|-\rangle_{\hat{v}}$.

To each unit vector $\hat{u}$ (i.e., to each point on the unit sphere), one can associate the projection operator
\begin{equation}
\label{projection operator}
P_{\hat{u}}=|+\rangle_{\hat{u}}\phantom{.}_{\hat{u}}\langle +|
\end{equation}
onto the set of states having spin ``up'' along direction $\hat{u}$. Therefore, according to Born rule, the probability that a spin measurement along direction $\hat{u}$ gives the outcome ``up'' (i.e., value $\hbar/2$), when the system is prepared in state $|+\rangle_{\hat{v}}$, is given by:
\begin{equation}
\label{probability up}
{\cal P}(\hat{v}\to\hat{u})= \phantom{.}_{\hat{v}}\langle +|P_{\hat{u}}|+\rangle_{\hat{v}}=|\phantom{.}_{\hat{u}}\langle +|+\rangle_{\hat{v}}|^2.
\end{equation}
To calculate this probability, one can reason as follows: if $\gamma = \arccos{(\hat{u}\cdot\hat{v})}$ is the angle between $\hat{v}$ and $\hat{u}$, i.e., the angle that one needs to rotate $\hat{v}$, along direction $\hat{n} = \hat{v}\times\hat{u}$, to reach $\hat{u}$, and if we denote by $R_{\hat{n}}(\gamma)=\exp(-i\gamma S_{\hat{n}}/\hbar)$ the corresponding rotation operator, with $S_{\hat{n}}=\vec{S}\cdot\hat{n}$, we have $|+\rangle_{\hat{v}} \propto R_{\hat{n}}(\gamma) |+\rangle_{\hat{u}}$. 

Then, exploiting the remarkable properties of the exponential function and of the spin $2\times 2$ matrices (see any good book of quantum mechanics), it is not difficult to show that the rotation operator $R_{\hat{n}}(\gamma)$ can be written in the simple form:
\begin{equation}
\label{rotation operator}
R_{\hat{n}}(\gamma)=\left(\cos{\frac{\gamma}{2}}\right)\mathbb{I} -i\left(\frac{2}{\hbar}\sin{\frac{\gamma}{2}}\right)S_{\hat{n}}.
\end{equation}
Therefore, considering that the unit vector $\hat{n}$ is orthogonal to $\hat{u}$, according to (\ref{x-y component equation}) we have $\phantom{.}_{\hat{u}}\langle +|S_{\hat{n}}|+\rangle_{\hat{u}}=0$, and one immediately finds that:
\begin{equation}
\label{probability up2}
{\cal P}(\hat{v}\to\hat{u})= \cos^2{\frac{\gamma}{2}},
\end{equation}
and of course, since $P_{\hat u}+ P_{-\hat u} = \mathbb{I}$, we also have
\begin{equation}
\label{probability down}
{\cal P}(\hat{v}\to-\hat{u})= \sin^2{\frac{\gamma}{2}}.
\end{equation}

As is well known, orthodox quantum mechanics doesn't explain the reason of probabilities (\ref{probability up2}) and (\ref{probability down}), i.e., where the indeterminism they subtend comes from and how the system is able to actualize outcomes that are only potential prior to the measurement. Of course, what is lacking in the above description, and more generally in the axiomatic formulation of orthodox quantum mechanics, is the measuring apparatus, i.e., a description of how the apparatus precisely interacts with the system during an idealized measurement. 

In the case of a spin measurement, the apparatus is for instance the one used in a typical Stern-Gerlach experiment, made of a magnetic field with a strong gradient (exerting a torque on the magnetic moment of the quantum entity), plus a detection screen used to reveal which beam deflection (upward or downward) has been actualized. Unfortunately, a direct observation of what exactly happens ``behind the scenes,'' when the spin quantum entity interacts (in a non-local way) with the apparatus and manifests its presence by means of a specific (upward or downward) spot on the screen, appears to be beyond our today (and perhaps  also our tomorrow) experimental abilities/possibilities. 

However, this doesn't mean we have to renounce to form a coherent picture of what's going on during the spin measurement, for instance by finding a meaningful structural analogy that could reveal us what could be the nature of the ``gaming apparatus'' in use. This is precisely what Aerts did when he invented his numerous quantum-machines, which are (idealized) macroscopic ordinary objects able to reproduce, among other things, the puzzling non-Kolmogorovian quantum probabilities~\footnote{Quantum probabilities are non-Komogorovian in the sense that they do not obey the basic axioms of classical probability theory, named after Andrey Kolmogorov~\cite{Kolmogorov}. In particular, they violate the so-called additivity axiom, as a consequence of the fact that in quantum mechanics one adds probability amplitudes, rather than the probabilities themselves.}.

\subsection{The spin quantum-machine (SQM)}
\label{The spin quantum-machine (SQM)}

Let us now describe the spin quantum-machine (SQM) and see how its functioning can simulate a spin-$1/2$ quantum measurement. The machine is very simple: it is constituted by a single point particle localized inside the shell of a three-dimensional (Euclidean) empty sphere of unit radius, the possible states of which correspond to the different places it can occupy on its internal surface.

\begin{figure}[!ht]
\centering
\includegraphics[scale =.6]{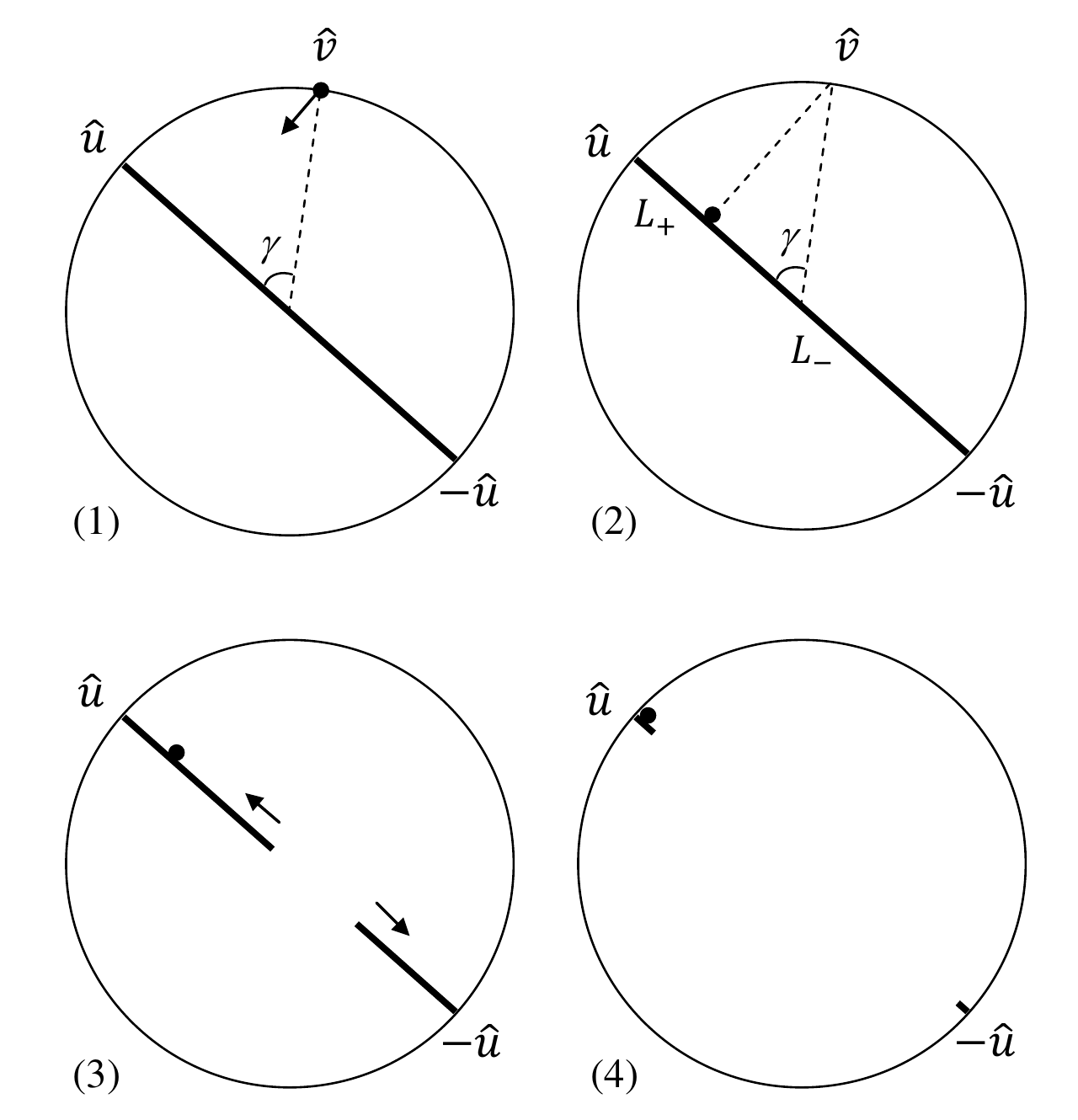}
\caption{A schematic representation of the spin quantum machine experiment, in the plane of the $3$-dimensional sphere where it takes place. (1) The elastic band is stretched and fixed at the two diametrically opposite end points $\hat{u}$ and $-\hat{u}$, then (2) the particle ``falls'' from its original place $\hat{v}$ onto the elastic band, taking the shortest path, then sticks to it, thus defining the two lenghts $L_+$ and $L_-$. (3) Finally, the elastic breaks, thus contracts, carrying with it the particle, which (4) is then pulled to one of the opposite points (here $\hat{u}$), defining its new state.\label{spin quantum machine}}
\end{figure}

Let us assume that the point particle, in a given moment, is located in the position specified by the unit vector $\hat{v}$. We introduce a specific class of experiments $e_{\hat{u}}$, that can be performed on it, which are so defined. To carry out experiment $e_{\hat{u}}$, the following procedure has to be executed: a uniform sticky elastic band is stripped between the two opposite points $\hat{u}$ and $-\hat{u}$, therefore passing through the centre of the sphere [Fig.~\ref{spin quantum machine}, picture (1)]. Once the sticky elastic band is placed, one lets the point particle ``fall'' from its original location (specified by $\hat{v}$) orthogonally onto the elastic, and stick to it, in a given point [Fig.~\ref{spin quantum machine}, picture (2)].

Then, one waits until the elastic breaks, at some unpredictable point (the rubber used to make the band is such that, once stretched, after a short time inevitably it breaks), and therefore the particle, which is attached to one of the two pieces of it, will be pulled to one of the opposite end points, $\hat{u}$ or $-\hat{u}$, thus producing the outcome of the experiment, i.e., the final position-state of the particle, acquired as a result of the $e_{\hat{u}}$-measurement [Fig.~\ref{spin quantum machine}, pictures (3) and (4)].

Of course, the experimenter cannot know in advance in which point the elastic will break, as this depends on a number of fluctuating factors which are totally beyond her/his possibility of control. However, s/he knows that the elastic is uniform and therefore can reasonably assume that the probability that it will break in a given segment is proportional to the length of that segment, and under this `natural' hypothesis, s/he can easily calculate the probabilities associated to the two mutually exclusive outcomes. 

Indeed, the probability that the particle ends up in point $\pm\hat{u}$ is given by the length $L_\pm$ of the piece of elastic between the particle and the end-point, divided by the total length of the elastic (which is twice the unit radius). Therefore, if $\gamma$ is the angle between $\hat{v}$ and $\hat{u}$, we obtain that the probability for outcome $\pm\hat{u}$ is given by:
\begin{equation}
\label{probabilities quantum machine}
{\cal P}(\hat{v}\to\pm\hat{u}) =\frac{1}{2}(1\pm\cos\gamma)=
\begin{cases}
\cos^2\frac{\gamma}{2} \\
\sin^2\frac{\gamma}{2}, 
\end{cases}
 \end{equation}
which exactly corresponds to the previously calculated quantum probabilities (\ref{probability up2}) and (\ref{probability down}), for measuring the spin $S_{\hat u}$ of a spin-$1/2$ quantum entity prepared in state $|+\rangle_{\hat{v}}$.

So, the spin quantum machine, which is only made of (idealized) classical entities, is perfectly able to replicate the behavior of a spin-$1/2$ entity, i.e., of  a pure quantum entity, and produce the same probability calculus.

\subsection{Discussion}
\label{Discussion}

The SQM model reveals a number of important features of quantum systems that we are now going to discuss, but before doing that, let us  point out an important difference between a spin-$1/2$ quantum entity and the point particle of the SQM. 

When we want to determine the spin state of a quantum entity, all we can do is to perform measurements, for instance by means of a Stern-Gerlach apparatus whose magnetic field gradient is oriented along a given direction $\hat{u}$. In the same way, in the SQM, one disposes of experiments $e_{\hat{u}}$, which are the equivalent of the Stern-Gerlach measurements. 

But in the SQM one also has the possibility of looking in the machine and directly ``see'' the state position of the point particle, as one can usually do, in principle, with whatever classical corpuscle. This possibility, of directly looking in the machine and discover, at any time, what's happening, is precisely what confers the SQM its strong explicative power, i.e., the possibility of fully visualizing the measurement process, as it evolves. 

However, such an ``insight'' should not be considered as a possibility to be used in practical terms, as an alternative to experiments $e_{\hat{u}}$, to determine the position-states of the point particle. As is the case with a spin-$1/2$ quantum entity, we must here do \emph{as if} the only experiments at our disposal to ``find out'' what are the locations of the point particle on the sphere, are the $e_{\hat{u}}$ ones, and no others. Having said that, let us now observe what we can learn from the structural analogy offered by the SQM. 

\emph{Observer effect}. A quantum measurement, in general, changes the state of the entity which is measured. This appears in very clear terms in the SQM: prior to the $e_{\hat{u}}$-experiment the point particle is in a position-state $\hat{v}$, whereas at the end of it its state is either $\hat{u}$ or $-\hat{u}$. This  means that a quantum measurement cannot be considered as a process of mere \emph{discovery} of properties that are already possessed by the system prior to the experiment, but as a process of \emph{creation} of those same properties that are measured. To quote Pascual Jordan~\cite{Jordan}: ``Observations not only disturb what has to be measured, they produce it [...]. We compel [the electron] to assume a definite position [...]. We ourselves produce the results of measurements.''

In other terms, the state (\ref{spin state}) cannot be interpreted as a description of the experimenter's lack of knowledge about whether the spin is ``up'' or ``down'' with respect to the $\hat{z}$ direction, as these ``up'' and ``down'' outcomes are literally created during the interaction of the entity with the experimental apparatus, and weren't existing prior to it. Of course, all this is already manifest in the quantum formalism, as is clear that states, represented by rays of the Hilbert space, are typically changed by a measurement into other states, which are eigen-rays of the operator corresponding to the measurement. However, in the SQM we can explicitly observe that it is the \emph{invasiveness} and \emph{unpredictability} of the interaction between the entity and the measuring apparatus that is responsible for the creation of a condition that wasn't actual before the experiment.

\emph{Non-commutativity}. The non-commutativity of certain quantum observables (in the present case of $S_{\hat{x}}, S_{\hat{y}}$ and $S_{\hat{z}}$) appears to be a direct consequence of the invasive character of quantum measurements. Indeed, if what is measured is not discovered but created by the measurement, then the order with which one performs two subsequent measurements is going to deeply affect the final outcome, in the same way as putting on a sock, then a shoe, is not the same as putting on a shoe and then a sock. 

In the SQM, non-commutativity is evidenced by the fact that, if one performs an experiment $e_{\hat{u}}$ (the equivalent in the machine model of a measurement of spin component $S_{\hat{u}}$), followed by an experiment $e_{\hat{w}}$, the particle's final position will be either $\hat{w}$ or $-\hat{w}$, whereas if one performs first $e_{\hat{w}}$ and then $e_{\hat{u}}$, the final position will be either $\hat{u}$ or $-\hat{u}$. Therefore, if $\hat{u}\neq\hat{w}$, these final states will necessarily be different, showing that the two experiments are incompatible, as their order of execution is relevant and cannot be commuted.

\emph{Hidden measurements}. The nature of the interaction between the point particle and the elastic in the SQM model reveals a possible fascinating feature of quantum measurements. Imagine for a moment that, instead of a uniformly breakable elastic band, one uses a band that can only break in a specific predetermined point $x$ (the sphere's origin corresponding to $x=0$). Given the initial position-state $\hat{v}$ of the point particle, it is clearly possible in this case to predict in advance if the outcome will be $\hat{u}$ or $-\hat{u}$. Indeed, if point $x=1$ corresponds to position $\hat{u}$ on the sphere, and $x=-1$ to position $-\hat{u}$, then for $x<\cos\gamma$ the outcome will certainly be $\hat{u}$, and $-\hat{u}$ for $x>\cos\gamma$.

An elastic band that breaks in a single predetermined point also constitutes an invasive experiment, altering the state of the point-particle, but it does so in a perfectly predictable way (excluding of course the special case $x=\cos\gamma$). Let us call $e_{\hat{u},x}$ such a deterministic experiment. If we envisage an experiment whose procedure consists in choosing randomly one of the $e_{\hat{u},x}, x\in[-1,1]$, then executing it (which means that $x$ is considered to be a uniformly distributed random variable), then, again, we can only predict the outcomes in probabilistic terms, and the probabilities will be exactly given by (\ref{probability up2}) and (\ref{probability down}).

This means that in the experiment $e_{\hat{u}}$ (i.e., in the uniformly breaking elastic band) a collection of deterministic (potential) experiments $e_{\hat{u},x}$ are in fact ``hidden,'' and the way in which one of these hidden deterministic experiments is selected during the measurement process is totally beyond the knowledge and possibility of control of the experimenter, as it depends on the presence of fluctuations in the experimental context. 

This situation has been called \emph{hidden measurements}, by analogy with  so-called \emph{hidden variables} theories, where the lack of knowledge is hypothesized being in relation to the state of the system. Now, in general terms, one can show that if a theory describes a situation where the lack of knowledge is about the state of the physical entity, then it is necessarily a classical statistical theory, obeying Kolmogorov's axioms. On the other hand, if a theory describes a situation where the lack of knowledge is about the measurement to be actually performed, and such a measurement affects the state of the physical entity, then it is a non-classical statistical theory, violating classical Kolmogorov's axioms, as quantum theory does~\cite{Aerts7, Aerts7-b}.

In other terms, what the spin quantum-machine strongly suggests, is that the fundamental difference between classical and quantum probabilities lies in the fact that the former describes a situation of lack of knowledge about the state of the system, whereas the latter describes a situations where there is knowledge of the state of the entity, but lack of knowledge about the interaction between the measurement apparatus and the entity.

\emph{Intermediate systems}. The spin quantum-machine is in fact a much richer model than what we have shown so far. Indeed, it is very easy to modify its functioning to obtain probabilistic models that generalize the classical and quantum ones. To do this, Aerts considered a generalized class of experiments, employing elastics of a more complex structure. Let us call them $e_{\hat u}(\epsilon)$, where $\epsilon$ is a real parameter comprised between $0$ and $1$ (this modified spin quantum-machine model is then called $\epsilon$\emph{-model}~\cite{Aerts3}). 

\begin{figure}[!ht]
\centering
\includegraphics[scale =.6]{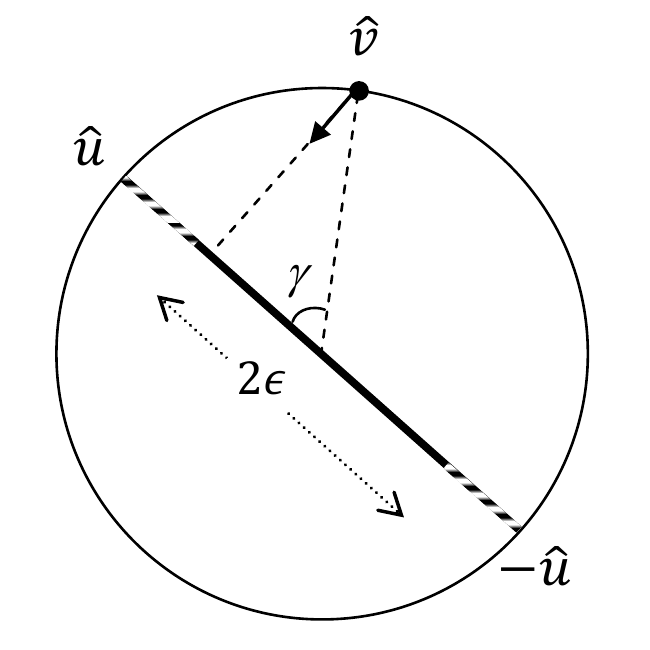}
\caption{A schematic representation of a $e_{\hat u}(\epsilon)$-measurement, using an elastic band that can only break in its central segment of length $2\epsilon$, whereas it is unbreakable in its lower and upper segments of length $\frac{1}{2}-\epsilon$.
\label{spin epsilon machine}}
\end{figure}

An $e_{\hat u}(\epsilon)$-measurement has the same protocol of a $e_{\hat u}$-measurement, but this time the elastic band used is not anymore uniformly breakable, but breakable only in a segment of length $2\epsilon$ around its center, and unbreakable in its lower and upper segments, as depicted in Fig.~\ref{spin epsilon machine} (we describe here a simplified version of the model, which was presented in~\cite{Aerts4b}).

Therefore, an $e_{\hat u}(1)$-experiment is a measurement using a fully uniformly breaking elastic band [$e_{\hat u}(1)=e_{\hat u}$] and corresponds to the pure quantum situation where there is \emph{maximum lack of knowledge} about the point where the elastic is going to break. This is the situation described by the SQM, that we have previously analyzed, whose probabilities are given by (\ref{probability up2}) and (\ref{probability down}). On the other hand, an $e_{\hat u}(0)$-experiment is a measurement using an elastic band that breaks exactly in its middle, with certainty [$e_{\hat u}(0)=e_{{\hat u},0}$] and corresponds to a purely classical (deterministic) situation with \emph{minimum lack of knowledge} about the point where the elastic is going to break. 

But what about a general $e_{\hat u}(\epsilon)$-experiment, with $0<\epsilon <1$, i.e., an experiment using an elastic band which can only (uniformly) break around its center, in a segment of length $2\epsilon$? The associated probabilities are easy to calculate, and one has to distinguish the following three cases:

(1) If the particle, when it ``falls'' orthogonally onto the elastic, lands on its upper unbreakable segment ($\hat{v}\cdot\hat{u}=\cos\gamma\geq\epsilon$), then:
\begin{equation}
\label{probabilities epsilon machine1}
{\cal P}^\epsilon(\hat{v}\to\pm\hat{u})=
\begin{cases}
1 \\
0, 
\end{cases}
 \end{equation}

(2) If the particle, when it ``falls'' orthogonally onto the elastic, lands on its central uniformly breakable segment of length $2\epsilon$, ($-\epsilon <\cos\gamma<\epsilon$), then:
\begin{equation}
\label{probabilities epsilon machine2}
{\cal P}^\epsilon(\hat{v}\to\pm\hat{u})=\frac{1}{2\epsilon}(\epsilon \pm\cos\gamma).
\end{equation}

(3) If the particle, when it ``falls'' orthogonally onto the elastic, lands on its lower unbreakable segment ($\cos\gamma\leq -\epsilon$), then:
\begin{equation}
\label{probabilities epsilon machine3}
{\cal P}^\epsilon(\hat{v}\to\pm\hat{u})=
\begin{cases}
0 \\
1. 
\end{cases}
 \end{equation}

To what kind of situation do the above probabilities correspond? Do they correspond to a classical or to a quantum probability model? In fact, to none of them: they don't fit into a classical Kolmogorovian probability model, but neither into a non-Kolmogorovian probability model of the quantum kind, as they truly correspond to new intermediate models, describing more general structures, of which the classical and quantum ones are limit cases~\cite{Aerts7,Aerts10,Aerts11}. 

What is interesting here to emphasize, in the general analysis of the SQM model and of its $\epsilon$-model generalization, is that our knowledge about the behavior of the physical entity is clouded by the presence of fluctuations, whose dynamics is beyond our control power. In the ambit of the $\epsilon$-model the amplitude of these fluctuations can be quantified by the real parameter $\epsilon\in[0,1]$. When $\epsilon =1$, the fluctuations which are responsible for the dispersion in the results of the measurement are maximal, as is maximal our lack of knowledge about the experimental outcome. This corresponds to the limit case of a pure quantum system. However, when $\epsilon$ decreases, but is not zero, we are in a situation of \emph{intermediate knowledge}: for certain preparations of the physical system we can predict with certainty the outcome, whereas for others we are still in a situation where they can only be predicted in strict probabilistic terms. 

Then, when $\epsilon$ reaches its lower value zero, we can say that the fluctuations in the measuring apparatus are in a sense smaller than the scale of the physical entity, and therefore cannot affect it in terms of outcomes (the different possible experiments that are randomly selected are not distinguishable in terms of their outcomes), so that again we are in a situation of maximal knowledge, typical of classical determinism. 

The above scheme describes a possible approach to the solution of the longstanding problem of finding a meaningful quantum-to-classical limit~\cite{Aerts-Durt, Aerts-Durt2}, as it provides the possibility of a continuous (non-singular) transition from these two limit situations.  

A last remark is due. Usually it is believed that macroscopic entities, like the ordinary entities with which we interact daily, cannot exhibit quantum (or intermediate, quantum-like) behaviors. As the SQM model clearly demonstrates, such a belief is unfounded. Of course, for the reader it may be difficult to accept that a point particle on a sphere can also be considered a quantum entity. However, one should not forget that its quantum character is revealed when we perform on it certain experiments, and not others. We must not forget that the $e_{\hat u}$ experiments are, in the SQM model, the only experiments that we are allowed, by definition, to execute. 

We are not allowed to ``see'' the point particle by other means, and this is the reason why, considering this restriction, the point particle can behave as a quantum entity. As Aerts and Durt emphasize~\cite{Aerts-Durt}: ``This is indeed exactly the situation that we encounter when we make investigations about quantum entities in the micro-world. We cannot `see' or `touch' these entities, and have only knowledge about them by means of the experiments that we can carry out on them.''

\section{A classical macroscopic entity violating Bell's inequality}
\label{A classical macroscopic entity violating Bell's inequalities}

\subsection{Bell's inequality}
\label{Bell's inequalities}

Before describing (our alternative version, based on elastics, of) Aerts' \emph{connected vessels of water} model~\cite{Aerts2, Aerts5, AertsBroekaert}, let us briefly recall what Bell's inequality is all about~\cite{Bell1, Bell0}. Bell was able to translate into a mathematical inequality certain general assumptions about physical systems, so that if these equalities are found to be experimentally violated, then at least one of the assumptions that are used to derive them must be wrong. We shall not prove here Bell's inequality, but simply recall its expression. 

Assume that, on a given physical entity, four different experiments can be performed. Let us call them $e^A_a$, $e^A_{a'}$, $e^B_b$ and $e^B_{b'}$. Let us also call $o^A_a$, $o^A_{a'}$, $o^B_b$ and $o^B_{b'}$ the associated outcomes, which are assumed can only take the values $+1$ or $-1$ (if the experiments are described by self-adjoint operators, as it is the case in quantum mechanics, then the outcomes correspond to their eigenvalues). 

A further hypothesis is that experiments $e^A_a$ and $e^A_{a'}$ can also be performed together with either of experiments $e^B_b$ and $e^B_{b'}$, thus defining the additional \emph{coincidence} experiments $e_{ab}^{AB}$,  $e_{ab'}^{AB}$, $e_{a'b}^{AB}$ and $e_{a'b'}^{AB}$. To each coincidence experiment $e_{cd}^{AB}$, $c\in\{a,a'\}$, $d\in\{b,b'\}$, one can associate the expectation value $E^{AB}_{cd}$ of the product of outcomes $o^A_c o^B_d$, by:
\begin{eqnarray}
\label{expectation value}
E^{AB}_{cd}&=&\sum {\cal P}_{cd}^{AB}(o^A_c,o^B_d) o^A_co^B_d\nonumber\\
&=& +{\cal P}_{cd}^{AB}(+1,+1) + {\cal P}_{cd}^{AB}(-1,-1)\nonumber\\
&\phantom{=}& - {\cal P}_{cd}^{AB}(+1,-1) - {\cal P}_{cd}^{AB}(-1,+1),
\end{eqnarray}
where ${\cal P}_{cd}^{AB}(o^A_c,o^B_d)$ is the probability that the coincidence experiment $e_{cd}^{AB}$ yields the outcomes $(o^A_c,o^A_d)$. 

Then, under the assumption that the outcomes are independently determined by some hidden variables, so that the expectation (\ref{expectation value}) can be written as the integral of the product of the two outcomes over these hidden variables (an assumption often referred to as \emph{Bell locality}), it is possible to prove the following (Bell) inequality~\cite{Bell1, Bell0}:
\begin{equation}
\label{Bell inequalities}
|E^{AB}_{ab} - E^{AB}_{ab'}| + |E^{AB}_{a'b'} + E^{AB}_{a'b}|\leq 2.
\end{equation}

Bell derived this inequality having in mind the physical system originally introduced by Bohm~\cite{Bohm}, of two entangled spin-$1/2$ quantum entities in a \emph{singlet (zero) spin state}, which is an \emph{entangled}  (non product) state: 
\begin{equation}
\label{singlet state}
|\psi_S\rangle = \frac{1}{\sqrt{2}}\left(|+\rangle^A_{\hat{u}}\otimes |-\rangle^B_{\hat{u}} - |-\rangle^A_{\hat{u}}\otimes |+\rangle^B_{\hat{u}} \right),
\end{equation}
where $|\pm\rangle^A_{\hat{u}}$ and $|\pm\rangle^B_{\hat{u}}$ describe ``up'' and ``down'' eigenstates of the spin operator $S_{\hat{u}}=\vec{S}\cdot\hat{u}$, along the $\hat{u}$-direction, in two separated (and arbitrarily distant) regions of space $A$ and $B$, respectively. 

\begin{figure}[!ht]
\centering
\includegraphics[scale =.9]{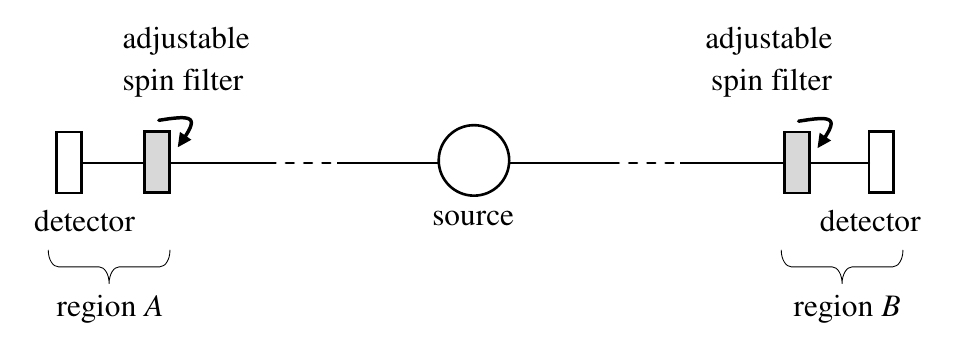}
\caption{A schematic representation of an experiment measuring the spin of two spin-$1/2$ entities in a singlet spin state, in two separated regions of space, $A$ and $B$. The spin filters, which can be independently oriented along different directions, can detect ``up'' and ``down'' spin components, by letting for instance pass the ``up'' spin components and stopping the ``down'' spin components.\label{coincidence}}
\end{figure}

On this singlet spin state, a certain number of spin measurements can be performed, using for instance two adjustable Stern-Gerlach filters, and the associated screen detectors, placed in regions $A$ and $B$, as schematically depicted in Fig.~\ref{coincidence}.

In this ambit, one can conceive the following four experiments: $e^A_a\equiv\frac{2}{\hbar}S_{\hat{a}}\otimes\mathbb{I}$, $e^A_{a'}\equiv\frac{2}{\hbar}S_{\hat{a}'}\otimes\mathbb{I}$, $e^B_b\equiv\mathbb{I}\otimes\frac{2}{\hbar}S_{\hat{b}}$ and $e^B_{b'}\equiv\mathbb{I}\otimes\frac{2}{\hbar}S_{\hat{b}'}$, where $\hat{a}$, $\hat{a}'$, $\hat{b}$ and $\hat{b}'$ are unit vectors describing the orientations of the Stern-Gerlach filters (orthogonal to the direction of flight of the pair of quantum entities forming the singlet state), $\mathbb{I}$ is the identity operator, $\otimes$ is the tensor product, and the symbol $\equiv$ is to be interpreted as ``the experiment on the left hand side corresponds to the measuring of the observable on the right hand side.'' 

These experiments (that can take values $+1$ or $-1$) can be combined to define the coincidence experiments:
\begin{equation}
\label{coincidence experiments quantum}
e_{cd}^{AB}\equiv\frac{4}{\hbar^2}S_{\hat{c}}\otimes S_{\hat{d}}, 
\end{equation}
where $\hat{c}\in\{\hat{a},\hat{a}'\}$, $\hat{d}\in\{\hat{b},\hat{b}'\}$.

Now, considering the quantum mechanical expectation value $E^{AB}_{cd}=\langle \psi_S|\frac{4}{\hbar^2}S_{\hat{c}}\otimes S_{\hat{d}}|\psi_S\rangle$, and using the properties of the spin operators (see any good book of quantum mechanics), one can show that $E^{AB}_{cd} = -\hat{c}\cdot\hat{d}$. Therefore, if one takes $\hat{a}$, $\hat{a}'$, $\hat{b}$ and $\hat{b}'$ to be coplanar and choose the angles between $\hat{a}$ and $\hat{b}$, $\hat{b}$ and $\hat{a}'$, and between $\hat{a}'$ and $\hat{b}'$ to be equal to $\pi/4$, one obtains:
\begin{eqnarray}
\label{Bell violation quantum}
\lefteqn{|E^{AB}_{ab} - E^{AB}_{ab'}| + |E^{AB}_{a'b'} + E^{AB}_{a'b}| =}\nonumber\\
& & \left|-\frac{\sqrt{2}}{2}+\left(-\frac{\sqrt{2}}{2}\right)\right|+\left|-\frac{\sqrt{2}}{2}-\frac{\sqrt{2}}{2}\right| =2\sqrt{2},
\end{eqnarray}
which is clearly a violation of (\ref{Bell inequalities}).

This result shows that no physical theory which is local in the sense specified by Bell, can agree with all the statistical implications of quantum mechanics. Bell locality hypothesis can be understood as an assumption about the \emph{experimental separation} of the two (here spin) entities emerging from the source and which, after an arbitrary long time of flight, are detectable in two regions $A$ and $B$ that are \emph{spatially separated} by an arbitrarily large distance. 

In accordance with Einstein's view, the assumption of many physicists was that a spatial separation would also imply an experimental separation, so that the expectations were that the microphysical reality would obey Bell's inequality. But these expectations were disregarded, as we today all know, by the historical experiments with entangled pairs of Aspect et al.~\cite{Aspect1, Aspect2} (who considered the equivalent situation of photons' polarization measurements, instead of spin measurements),  and the many others that since then followed, who confirmed the reality of the violation (\ref{Bell violation quantum}), and therefore the correctness of the quantum formalism describing entangled states.

In other terms, we have to renounce to local realism ``\`a la Bell'' in the description of the microworld. But, how can we \emph{understand} then the origin of the violation of Bell's inequality and, can they be violated only by microscopic quantum entities? As we shall see in the next section, Bell's inequality is easily violated also by macroscopic `classical' entities, provided we experiment on them in a very specific way. This will offer us the opportunity to see what exactly happens when a definite value of a property of the system in region $A$ is acquired by virtue of a measurement carried out in region $B$, and vice versa.

\subsection{An elastic band violating Bell's inequality}
\label{An elastic band violating Bell's inequality}

Let us now describe a macroscopic entity that can maximally violate Bell's inequality (\ref{Bell inequalities}). As we have mentioned in the Introduction, Aerts considered a system made of vessels of water connected by tubes and siphons~\cite{Aerts2, Aerts5, AertsBroekaert}. Here we shall present a much simpler system, which can play the same role as the connected vessels: a single uniform elastic band.\footnote{A very similar elastic band-system has also been utilized by Sven Aerts to obtain a realistic simulation of a so-called non-local PR box without communication.~\cite{Sven}}

We assume that the uniform elastic band has been made with a \emph{red} rubber and that when unstreched, its length is $L$. Two scientists are placed at the two ends of it (let us call them $A$ and $B$, respectively), and can perform certain experiments. Scientist $A$ can perform on the left end of the elastic band two experiments, $e^A_a$ and $e^A_{a'}$, which are defined as follow. 

\begin{figure}[!ht]
\centering
\includegraphics[scale =.9]{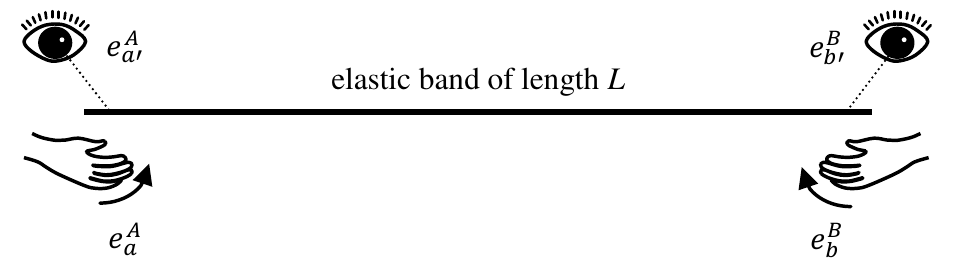}
\caption{A schematic representation of the experiments that scientists $A$ and $B$ can perform on the two distant ends of the elastic band entity, simultaneously or not, using their hands and eyes.\label{coincidence with elastic}}
\end{figure}

Experiment $e^A_a$ consists in grabbing the end of the elastic band and pulling it with force, then measuring the length of the unstreched elastic band that has been collected in this way: if it is greater than $L/2$, the outcome $o^A_a=+1$, otherwise $o^A_a=-1$. Experiment $e^A_{a'}$ is much simpler as it consists in simply looking at the elastic band and checking whether its color is red. If it is so, then the outcome $o^A_{a'}=+1$, otherwise $o^A_{a'}=-1$. Scientist $B$ can perform the same experiments as scientist $A$, but on the right end of the uniform elastic band. In other terms, $e^B_b$ is defined as $e^B_a$ and $e^B_{b'}$ as $e^B_{a'}$ (see the schematic representation of Fig.~\ref{coincidence with elastic}). 

Of course, since the elastic band is made with a red rubber and its length is $L$, all of the four above mentioned experiments, when performed individually, can only produce the outcome $+1$ (as is clear that when only one experimenter is pulling the elastic, the whole of it will come in her/his hands). The same remains obviously true when the coincidence experiments $e_{ab'}^{AB}$, $e_{a'b}^{AB}$ and $e_{a'b'}^{AB}$ are performed at the same time by the two scientists, whose outcomes are always $(+1,+1)$ (we observe that when the elastic is pulled by one of the experimenters, the other one still has the possibility to locally check its color, as is clear that the light coming from the elastic is always available to be detected at her/his location).   

The situation changes however when one considers the coincidence experiment $e_{ab}^{AB}$, which creates correlations. Indeed, if the two scientists grab and pull the elastic band at the same time, it will break in some unpredictable point, so that if one scientist collects a fragment of length greater than $L/2$, the other one will necessarily collect a fragment of length lower than $L/2$. Thus, the only possible outcomes of the coincidence experiment $e_{ab}^{AB}$ are $(+1,-1)$ and $(-1,+1)$, and of course they have the same probability to occur, which is equal to $1/2$.

According to (\ref{expectation value}), we therefore obtain that $E^{AB}_{ab'}= E^{AB}_{a'b}=E^{AB}_{a'b'}=1$, and $E^{AB}_{ab}=-1$, so that:
\begin{eqnarray}
\label{Bell violation classical}
\lefteqn{|E^{AB}_{ab} - E^{AB}_{ab'}| + |E^{AB}_{a'b'} + E^{AB}_{a'b}| =}\nonumber\\
& & \left|-1-\left(+1\right)\right|+\left|+1+\left(+1\right)\right| =4.
\end{eqnarray}

In other terms, not only the elastic band can be used to break Bell's inequality, but it can do it much more than the typical example of two coupled spin-$1/2$ entities in a singlet state.

\subsection{Discussion}
\label{Discussion2}

In~\cite{Mermin} David Mermin mentions a conversation with a distinguished physicist in Princeton. He asked him how he thought Einstein would have reacted to the violation of Bell's inequalities: ``He said that Einstein would have gone home and thought about it hard for several weeks that he couldn't guess what he would then have said, except that it would have been extremely interesting.''

This author likes to think that Einstein might have tried to find out an example ``\`a la Aerts,'' like his \emph{connected vessels model}, or the equivalent \emph{elastic band model} that we have described. And based on an example of this kind, he would have probably reached the same conclusions as Aerts did, regarding the real mystery hidden in Bell's inequalities (a.k.a. Bell's no-go theorem). 

What's immediately evident in the elastic band model is the fact that the entity playing the role of the quantum singlet state -- the uniform elastic band -- is a single genuine whole entity, and not two entities, and certainly cannot be understood as the sum of two separated interacting parts, each one described by its own specific state. 

Indeed, the elastic band is present not only in the two regions $A$ and $B$, where the two scientists are located (close to its ends), but also in between these two regions: it possesses what has been called the property of \emph{macroscopic wholeness}~\cite{Aerts2}, i.e., the property of \emph{hanging together through space}, which means it cannot be localized in different macroscopically separated spatial regions without also being localized somewhere in the region `in between' them.

This wholeness aspect was already pointed out by Schroedinger~\cite{Schroedinger}, when he emphasized that entangled states like (\ref{singlet state}) are associated to a notion of \emph{non-separability}, in the sense that two quantum entities can find each other in a state such that only the properties of the pair appear to be defined, whereas the individual properties of each one of the two entities that have formed the pair remain totally undefined~\footnote{In fact, and quite surprisingly, it is not the notion of \emph{non-separability} that creates difficulty in quantum mechanics, but rather the notion of \emph{separability}. In this regard, it is worth mentioning that the experimental violation of Bell's inequality didn't provide a solution of the EPR paradox per se. Such a solution only became possible in the eighties, following Aerts' deep analysis of the EPR situation, which he presented in his Doctoral Thesis and then further developed in a number of publications. In these works, he pointed out a very subtle and crucial point: that standard quantum mechanics is in fact unable, structurally speaking, to describe systems made of separated entities~\cite{Aerts5} (the concept of \emph{separation} is here to be understood in the experimental sense, i.e., in the sense that performing experiments on one of the entities doesn't affect the state of the others, and vice versa). As stressed by Aerts, this is a fundamental (and usually misunderstood) point, as EPR, in their (ex-absurdum) reasoning, precisely assumed, on the contrary, that quantum mechanics was perfectly able to describe the situation of two entities that become separated (in the experimental sense) as they fly apart in space. But this premise was proven by Aerts' analysis to be false. So, although EPR conclusion about the incompleteness of quantum mechanics was correct, it wasn't for the reasons advocated in their 1935 paper. If quantum mechanics is incomplete, it is because it fails to correctly describe experimentally separated physical entities~\cite{Aerts5}.}.

Clearly, the notion of non-separability introduced by Schroedinger becomes fully explicit and viewable with the elastic band example. Indeed, one can certainly confer well defined properties to the two distant ends of the unbroken elastic band, like their spatial distance, but on the other hand we cannot attribute to each of these distant ends, considered ``separately,'' well-defined properties, as they don't possess any independent reality outside of the reality of the whole band.

But when the coincidence experiment $e_{ab}^{AB}$ is executed, something very special happens: two entities are \textit{created} and correlations that weren't present before the experiment are in this way actualized. This creation of correlations, that can violate Bell's locality hypothesis (and therefore Bell's inequality), cannot be described in terms of local hidden variables, associated to the individual states of the (only potentially existing) separated fragments of the elastic band, as the correlations are produced by the measurement process itself, which in the present case is a \emph{hidden measurement} process.

As we observed for the SQM model, a hidden measurement process is a process characterized by two ingredients: (1) an aspect of creation of new properties and (2) an aspect of indeterminism, i.e., the fact that the outcome of such a creation process cannot be predicted in advance. Therefore, a natural question arises: is it the creation of correlations that were inexistent before the experiment that is responsible for the violation of Bell's inequality, or is it the indeterminism that is inherent in this process, or is it both?

This question was deeply analyzed in \cite{AertsBroekaert}, using a much more complex macroscopic model than the elastic band model presented here, able to violate Bell's inequality not in a maximal way, but exactly in the same way a quantum singlet state does~\cite{Aerts-a}. By varying in the model two parameters, quantifying respectively the degree of correlation and of indeterminism present in the system, the authors were able to show that the crucial ingredient in the violation of Bell's inequality is not the indeterminism, but the correlation, i.e., that it is the non-local aspect expressed by the correlation that is the true source of the violation, and that the presence of indeterminism can in fact decrease the value Bell's inequality takes. 

Without engaging in the rather involved analysis of the more complex situation presented in~\cite{AertsBroekaert}, let us however show, in our simpler elastic band model, how Bell's inequality can be violated even when all source of indeterminism in the experiments has been eliminated. 

To do this, we simply have to replace the hidden measurement promoted by the uniform elastic band, by a ``pure'' (non-hidden) measurement promoted by a (red) elastic band that can only break (if the elastic is stretched with sufficient force) in a predetermined point, located, say, at distance $L/3$ from its left end. 

Then, the outcomes of the coincidence experiments $e_{ab'}^{AB}$, $e_{a'b}^{AB}$ and $e_{a'b'}^{AB}$ will be again $(+1,+1)$, as it was the case for the uniform elastic band (scientists cannot break the elastic if they pull it only from one end). On the other hand, concerning experiment $e_{ab}^{AB}$, the only difference is that now we can predict in advance what will be the outcome, meaning that the correlation is created in a perfectly deterministic way, always producing the outcome $(-1,+1)$. 

Hence, as for the uniform elastic band, we find that $E^{AB}_{ab'}= E^{AB}_{a'b}=E^{AB}_{a'b'}=1$, and $E^{AB}_{ab}=-1$, so that (\ref{Bell violation classical}) holds and Bell's inequality is again maximally violated. This shows that it is the \emph{creation of correlations} which is really responsible for the violation, and not its deterministic or indeterministic character.  

Correlations that were not present before the experiment, but are created by and during it, are called \emph{correlations of the second kind}~\cite{Aerts2}. This to distinguish them from so-called \emph{correlations of the first kind}, which are already present before the experiment is executed, and are therefore not created but only discovered by it.  

Let us observe that although in the elastic experiment the correlation (of the second kind) is created instantly, in the precise moment when elastic breaks, the information relative to the correlation cannot travel faster than the speed of contraction of the elastic. In other terms, the elastic can't be used to produce superluminal signaling and Einstein locality is obviously preserved.~\footnote{This of course could also be considered as a weakness of the elastic-model, as is clear that it cannot simulate an instantaneous co-creation of information by the two experimenters, as it appears to happen instead in Aspect's experiments with entangled photons. But being the connection between entangled microscopic entities of a non-spatial nature (see the discussion in the last section of this paper), the times and ways through which the co-created information reaches the two experimenters needs not be limited by relativistic constraints, as is the case instead for macroscopic entities like an elastic band. Also, let us not forget that the most interesting aspect of the model is not its ability to perfectly simulate a specific microscopic quantum system, but to show that simple physical systems exist which are able to create correlations without communication, thus violating Bell's inequality (see also the discussion in~\cite{Sven}).}

\begin{figure}[!ht]
\centering
\includegraphics[scale =.9]{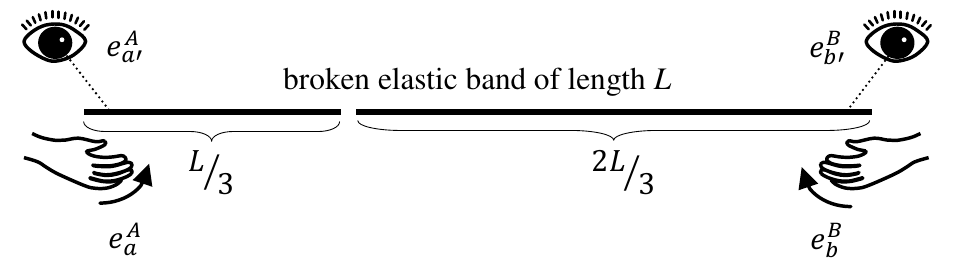}
\caption{A schematic representation of the experiments that scientists $A$ and $B$ can perform on the two distant ends of an \emph{already broken} elastic band entity, simultaneously or not, using their hands and eyes.\label{coincidence with broken elastic}}
\end{figure}

Let us also observe that, as explained in~\cite{Aerts2}, one cannot use correlations of the first kind to violate Bell's inequality. Let us show this explicitly in our model by replacing the whole elastic entity by an already broken elastic entity, where the left fragment is, say, of length $L/3$ and the right fragment of length $2L/3$ (see Fig.~\ref{coincidence with broken elastic}). 

Clearly, the respective lengths of the two elastic fragments are correlated, as is clear that their sum must be equal to the total length $L$ of the (previously) unbroken elastic band. However, this correlation is now of the first kind, as it is already present in the system before the experiments are carried out. 

Also in this case, the outcomes of all the experiments are predetermined, as it was the case for the (unbroken) elastic band only breakable in one point, and we have that the outcomes of $e_{a'b}^{AB}$ and $e_{a'b'}^{AB}$ are $(+1,+1)$, and the outcome of $e_{ab}^{AB}$ is $(-1,+1)$.  On the other hand, this time experiment $e_{ab'}^{AB}$ has outcome $(-1,+1)$, so that $E^{AB}_{a'b}=E^{AB}_{a'b'}=1$, and $E^{AB}_{ab}=E^{AB}_{ab'}=-1$, yielding:
\begin{eqnarray}
\label{Bell non-violation classical}
\lefteqn{|E^{AB}_{ab} - E^{AB}_{ab'}| + |E^{AB}_{a'b'} + E^{AB}_{a'b}| =}\nonumber\\
& & \left|-1-\left(-1\right)\right|+\left|+1+\left(+1\right)\right| =2,
\end{eqnarray}  
in plain accordance with Bell's inequality (\ref{Bell inequalities}).

We can therefore conclude with Aerts that~\cite{Aerts2}: ``The violation of Bell inequalities is not a property of micro-entities. Bell inequalities can equally well be violated by coincidence experiments on classical microscopic entity. In fact Bell inequalities can always be violated if during the coincidence experiments one breaks one entity into separated pieces, and by this act creates the correlations.''

We call attention to the fact that, because of its connectedness, a whole elastic band is the expression of a number of relations between its different parts. Each point of the elastic is indeed representative of a specific relation between the two non-separated parts of the elastic which are on the left and right sides of the point (the relation being expressed by the fact that the sum of the unstreched lengths of these two parts has to be $L$). 

Hence, one may be tempted to conclude that the violation of Bell's inequality is in fact the result of correlations of the first kind, between \emph{non-separated parts}, and not of correlations of the second kind, between \emph{separated fragments}. This however would be a wrong conclusion, as only correlations between separated fragments are observed during the experiments, and these correlations are literally created by them. Indeed, the non-separated parts of an unbroken elastic band only correspond to \emph{potential separated fragments}, and therefore to \emph{potential correlations}, when considered in relation to the outcomes of the experiments.~\footnote{To make this point even clearer, consider holding a die. All its six faces exist, but only when the die is rolled an \emph{upper face} is created by the ``rolling experiment.'' If we stick together two dice, we obtain a connected structure, exhibiting specific relations between the different faces of the double-die. All these relations exist, but it is only when the double-die is rolled that a specific couple of correlated \emph{upper faces} is created, a process which therefore has to be understood as a correlation of the second kind.}

This remark is important in order to understand that entangled states, like singlet states, are the expression of a (hidden) connected structure of relations out of which correlations between outcomes are created, not discovered. This is so because the properties which are observed during the coincidence experiments do not exist prior to them, but are actualized by them, in the same way as separated fragments do not exist until the whole elastic is broken. 

This mechanism of creation of correlations is already manifest in the quantum formalism, as is clear that the singlet state is a state of zero spin, which is rotational invariant, so that there is nothing special about the direction $\hat{u}$ we have chosen to represent the state in terms of ``up'' and ``down'' components: if one chooses another arbitrary direction, say $\hat{w}$, then by a direct algebraic calculation one can show that (\ref{singlet state}) can be equivalently written as:
\begin{equation}
\label{singlet state bis}
|\psi_S\rangle\propto \frac{1}{\sqrt{2}}\left(|+\rangle^A_{\hat{w}}\otimes |-\rangle^B_{\hat{w}} - |-\rangle^A_{\hat{w}}\otimes |+\rangle^B_{\hat{w}}\right),
\end{equation}

This means that the correlation between the outcomes of the left (region $A$) and right (region $B$) spin-measurements doesn't depend on the global  orientation chosen for the two spin filters (only their relative orientation is important), and this also means that $|\psi_S\rangle$ doesn't describe the state of two-entities having already actualized their spin, although the way it is mathematically written may wrongly suggest so. 

In fact, their spins are literally \emph{created} by the coincidence experiment, which in a sense ``breaks'' the singlet state into a state describing two separated spin entities with a specific direction (although now entangled with the experimental apparatus), exactly in the same way an elastic band of length $L$ doesn't describe two actually separated elastic bands, but only two potentially separated elastic bands, which can be created (in a deterministic or indeterministic way) by performing a correlation experiment of the breaking kind.

\section{Concluding remarks}
\label{Concluding remarks}

Let us briefly summarize the main results we have presented in this paper. To understand what could possibly happen ``behind the scenes,'' during a quantum measurement, we have reviewed two of Aerts' macroscopic models: the SQM (and its $\epsilon$-model generalization)~\cite{Aerts3, Aerts3-a}, and an elastic band model inspired by a previous connected vessels of water model~\cite{Aerts2, Aerts5, AertsBroekaert}. 

In this way, we have gained considerable insight into the functioning of the quantum level of our reality, which therefore appears to be less mysterious than expected, considering that also ordinary macroscopic entities can give rise to quantum probabilities, or violate Bell's inequalities, and that the way they do it is perfectly under our eyes. 

More precisely, thanks to the SQM model, and the structural analogies it provides, we have explained the emergence of quantum probabilities in terms of hidden, invasive measurements, which are selected in a way that cannot be controlled and therefore predicted by the experimenter. 

This of course doesn't mean that we now understand everything about quantum measurements, which surely retain part of their mystery. Indeed, the SQM model just tells us in what direction one should search, but not what exactly to find: we have to search not for hidden variables associated to the state of the entity, as it has been historically done, but for hidden variables associated to the measurement process, i.e., for the ``pure'' (possibly deterministic) measurement interactions which are selected through a (symmetry breaking) mechanism that cannot be controlled (and therefore be known) by the experimenter.

With regard to the nature of these hidden measurements, as far as the present author can judge, the mystery remains intact. Of course, there exist unconventional approaches to quantum mechanics where some attempts have been made to identify possible (real) processes that would operate at a subquantum level, and that could possibly explain the emergence of quantum probabilities. 

Since we have quoted John Cramer in the Introduction, let us mention, as a simple paradigmatic example, his \emph{transactional interpretation} (TI)~\cite{Cramer}, which has been recently revisited by Ruth Kastner~\cite{Kastner}. TI is a time-symmetric re-interpretation of conventional quantum mechanics, such that kets are associated to retarded ``offer waves of possibilities,'' evolving forward in time, emitted by entities which, in a given context, play the role of sources, whereas bras are associated to advanced ``confirmation waves of possibilities,'' evolving backward in time, emitted by entities playing the role of absorbers, in response to the offers received from the sources. These a-temporal offer-confirmation exchanges between sources and potential absorbers, give rise to so-called \emph{incipient transactions}, which are then selected (through a symmetry breaking mechanism) to give rise to \emph{actualized transactions}, i.e., to processes that select specific outcomes in the form of actual transfers of given quanta of energy and momentum. 

It is not our intention to enter here into the details of the TI and discuss the strengths and weaknesses of this rather unusual interpretation. Our only intention was to provide a very simple example of an approach whose ontology contemplates a candidate for something like a hidden measurement interaction, which in the TI case would correspond to the process of actualization of incipient transactions.

Regarding the hidden-measurement approach, some readers may be tempted to believe that, because of the well-known Gleason's theorem~\cite{Gleason} and  Kochen-Specker's impossibility proof~\cite{Kochen}, it would be unfeasible to construct models like the SQM, for Hilbert spaces of dimension greater than 2. This is however not the case, as No-Go theorems for hidden variables only apply to models with hidden variables referring to the \emph{state of the system}, and not to models where the hidden variables refer instead to the \emph{measurement process} (see the discussion in~\cite{Aerts-3d} and the references cited therein).

In other terms, although one could  consider, in light of the above results, that the SQM model would not per se be sufficient to possibly explain the origin of quantum probabilities, one should bear in mind that its built-in hidden-measurement mechanism can be easily generalized to describe the probability model of quantum systems of arbitrary dimension, as it was done by Aerts in~\cite{Aerts7}. A discussion of these more general situations would however go beyond the scope of the present paper, and we refer the interested readers to~\cite{Aerts4,Aerts4b,Aerts7,Aerts10,Aerts-3d}, and the references therein.

Let us now comment on the second part of the present paper: the violation of Bell's inequalities. Thanks to the elastic band model, we can easily affirm that much of the strangeness associated to these phenomena, which disturbed so much Einstein, Schroedinger, Bell and many others, is gone. Indeed, it is possible to explain quantum correlations, the ``spooky actions at a distance,'' as Einstein used to call them, as processes during which a whole entity is broken into parts, in a deterministic or indeterministic way. 

Again, this doesn't mean we now understand everything about the quantum correlations produced by entangled states. Indeed, as we explained, for a macroscopic entity to be able to violate Bell's inequality, the \emph{conditio sine qua non} is that it possesses the property of \emph{macroscopic wholeness}: it has to be present in the spatially separated regions $A$ and $B$, where the coincidence experiments are performed, but it has also to be present in the region between them. 

But quantum microscopic entities do not possess this macroscopic wholeness property, and this certainly constitutes that part of the mystery that remains intact when considering the violation of Bell's inequality at the microscopic level, in EPR-like experiments. Indeed, considering a singlet state formed from a pair of electrons that have flown apart, and have reached the distant regions $A$ and $B$, respectively, we know that one electron can be detected with probability close to one in region $A$, and same for the other electron in region $B$, but that the probability of detecting them in between is almost zero. So, it seems that at the microscopic level, spatially separated entities can nevertheless remain interconnected, not through space, but in some other, yet to be explained, way. 

Of course, the difficulty many physicists experience in truly understanding this possibility lies once more in the difficulty one might have in visualizing a concrete model that would render it manifest. This however is not unfeasible. A possibility has been proposed in~\cite{Sassolimachine}, where a generalization of the concept of macroscopic wholeness was proposed, called \emph{process-macroscopic wholeness}, expressing the idea that two entities can remain connected not only through space, in a static way, but also \emph{through time}, in a more dynamic way.

These kind of explanations force us to abandon an old preconception, which may be the one truly hindering our understanding of the microworld:  that microscopic entities should always be present in our three-dimensional space and that, more generally, reality should only exist within space. On the contrary, it seems reasonable to hypothesize that microscopic entities are non-spatial entities, and that, quoting Aerts, space would only be~\cite{Aerts4}``[...] a momentaneous crystallization of a theatre for reality where the motions and interactions of the macroscopic material and energetic entities take place. But other entities - like quantum entities for example - `take place' outside space, or - and this would be another way of saying the same thing - within a space that is not the three dimensional Euclidean space.''

Clearly, we cannot enter here in the analysis of these important ideas (for more details, see for instance~\cite{Aerts2,Aerts3,Aerts4,Massimiliano1,Massimiliano3,Sassolimachine} and the references cited therein), as this would bring us too far away from the primary motivation of this paper, which was only to conceptually review two of Aerts' historical machine-models, in order to promote a deeper understanding of the physical content ``hidden'' within the abstract quantum formalism. It is worth emphasizing that the truly interesting aspect about these models is not their ability to realistically describe physical entities as such, but to capture, by means of powerful structural analogies, the possible logic at the basis of their functioning, when they interact with the different experimental contexts.

Also, these machine-models force us to abandon too simplistic distinctions about classical and quantum entities. Indeed, the quantum behavior of a physical entity appears to also depend on how we choose to actively experimenting on it, according to specific protocols. And this is the reason why also macroscopic ordinary entities can behave in a quantum-like fashion. This remark has some relevance also in relation to possible \emph{loopholes} in experiments testing Bell's inequality, where by loophole we mean here a more or less consciously used assumption which would undermine the validity of an experiment (see for instance~\cite{Auletta}, pages 600-605). 

The analysis here presented highlights one of these possible loophole: the (wrong) assumption that Bell's inequality could only be violated by microscopic systems and that the violation would be an indication of the presence of only microscopic quantum properties. On the contrary, we know that, quoting once more Aerts,~\cite{AertsBroekaert} ``Bell inequalities can be genuinely violated in situations that do not pertain to the microworld. Of course, this does not decrease the peculiarity of the quantum mechanical violation in the EPRB experiment. What it does, is shed light on the possible underlying mechanisms and provide evidence that the phenomenon is much more general than has been assumed.''

In other words, also the detecting apparatus, in a typical coincidence experiment, when considered as an interconnected macroscopic whole entity, is in principle able to produce correlations of the second kind, which could  be in principle at the origin of the observed violation of Bell's inequality. This possibility is sometimes called the ``locality loophole.''~\cite{Auletta} To close this loophole, one has to design the coincidence experiments in such a way to be sure that only the correlations originating from the entangled microscopic entity are at the origin of the violation of Bell's inequality. 

For this, one has to exploit the fact that, because of the non-spatial nature of microscopic entities, the effects resulting from the creation of correlations by entangled states ``travel'' virtually instantaneously (certainly thousands of times faster than the light speed~\cite{Salart}, although this cannot be used to produce superluminal signaling~\cite{Eberhard}), whereas possible effects due to correlations resulting from the interconnectedness of the detectors certainly cannot be communicated at speeds greater than the speed of light.  

This was done by Aspect, in his famous polarization experiments with entangled photons (see~\cite{Aspect1, Aspect2} and the references therein), where the choice of orientations of the polarizers were changed randomly in very short times, during the flight of the two entangled photons, in order to dramatically enforce relativistic separation. Thanks to these experiments, it was possible to conclude, quoting Aspect,~\cite{Aspect2} ``[...] that an entangled EPR photon pair is a non-separable object; that is, it is impossible to assign individual local properties (local physical reality) to each photon. In some sense, both photons keep in contact through space and time.''

To conclude, we point out that many other machine models have been invented in addition to those mentioned in this article, and we can only invite the reader to take cognizance of them, to get advantage of the numerous insights they are able to promote. Let us mention Aerts' model describing subsystems connected by rigid rods, which can violate Bell's inequalities in a non-maximal way, as microscopic quantum entities can do~\cite{AertsBroekaert, Aerts8-b}, also revealing possible unexpected interpretations of mixed states in the standard quantum formalism~\cite{Aerts8}. And let us also mention a recent model that has been presented by this author, called the $\delta$-quantum machine, which is able to replicate the scattering probabilities of a one-dimensional quantum scattering process by a Dirac delta-function potential. 

But Aerts, in his attempt to truly understand quantum mechanics, has not only invented mechanistic models. In more recent years he has also devoted a considerable effort in using the quantum mechanical formalism to model, with great success, human conceptual situations as they appear in cognition, decision theory and economics. This led him to ask a deep and thought provoking question~\cite{Aerts15}: ``If quantum mechanics as a formalism models human concepts so well, perhaps this indicates that quantum particles themselves are conceptual entities?'' 

This question was the starting point of a new interpretation of quantum mechanics, which is probably today's most advanced explanatory framework to understand this puzzling theory. According to Aerts, quantum entities would~\cite{Aerts15} ``[...] interact with ordinary matter, nuclei, atoms, molecules, macroscopic material entities, measuring apparatus, ..., in a similar way to how human concepts interact with memory structures, human minds or artificial memories.'' Therefore, quoting again Aerts~\cite{Aerts15}: ``if proven correct, this new quantum interpretation would provide an explanation according to which `quantum particles' behave like something we are all very familiar with, and have direct experience with, namely concepts.'' 

We will certainly not comment any further this subtle explanatory framework and its effectiveness in truly explaining phenomena such as entanglement and non-locality, which are traditionally considered (in the spirit of Feynman's quote cited in the Introduction) as ``not understood,'' and leave to the reader the intellectual pleasure to directly discover them in Aerts' recently published articles~\cite{Aerts15, Aerts16}. Let us just observe that it is rather remarkable that what is probably the best available model to understand the mysteries of the quantum level of our reality has always been at hand, hidden in the very structure of our cognitive processes.

\begin{acknowledgements}
I'm pleased to dedicate this article to Diederik Aerts, a scientist of our times who, like the giants of the past, is promoting through his multi-faceted research the possibility of reaching a true understanding of the working of our mysterious reality.
\end{acknowledgements}

\end{document}